\newcommand{\bq}{\begin{equation}}
\newcommand{\eq}{\end{equation}}
\newcommand{\bqa}{\begin{eqnarray}}
\newcommand{\eqa}{\end{eqnarray}}

\newcommand{\f}{\varphi}
\newcommand{\h}{{1\over2}}

\newcommand{\eps}{\varepsilon}

\documentclass[a4paper]{article}

\usepackage{epsfig}
\usepackage{axodraw}
\usepackage{graphicx}
\usepackage{amsmath}
\usepackage{makeidx}
\usepackage[top=1in, bottom=1in, left=1in, right=1in]{geometry}
\usepackage{cite}

\begin{document}

\begin{center}
\LARGE
Quantum Extremism: Effective Potential and Extremal Paths\\[30pt]
\large
E.N. Argyres\\[5pt]
Institute of Nuclear Physics, NCSR "Demokritos", Athens, Greece\\[20pt]
M.T.M. van Kessel\footnote{M.vanKessel@science.ru.nl} \hspace{50pt} R.H.P. Kleiss\footnote{R.Kleiss@science.ru.nl}\\[5pt]
IMAPP, FNWI, Radboud Universiteit Nijmegen, Nijmegen, The Netherlands\\[20pt]
July 7, 2009\\[50pt]
\end{center}

\section{Abstract}

The reality and convexity of the effective potential in quantum field theories has been studied extensively in the context of Euclidean space-time.  It has been shown that canonical and path-integral approaches may yield different results, thus resolving the `convexity problem'. We discuss the transferral of these treatments to Minkowskian space-time, which also necessitates a careful discussion of precisely which field configurations give  the dominant contributions to the path integral. In particular, we study the effective potential for the $N=1$ linear sigma model.

\newpage

\section{Introduction}

In general literature on QFT (in Minkowski space-time) it is stated that an effective potential is always \emph{real} and \emph{convex} \cite{Peskin, Weinberg, Riversbook, Symanzik, Iliopoulos, Haymaker, ORaifeartaigh, Fujimoto, Callaway, Bender, Cooper, Hindmarsh, Rivers, Wipf, Fukuda, Ringwald, Branchina, Wu, Dannenberg, Wiedemann}. This statement seems to have its origin in Euclidean space-time, where it can be proven rigorously \cite{Riversbook, Haymaker, Callaway, vanKessel, vanKesselthesis}. After having proven convexity for the effective potential (and even for the whole effective action) in the Euclidean case one then \emph{assumes} that the proof can be continued to Minkowski space-time.

In Euclidean space-time QFT there occurs the so-called convexity problem \cite{Peskin, Fujimoto, Bender, Cooper, Dannenberg, vanKessel, vanKesselthesis}. When one computes the effective potential in the canonical (perturbative) way one finds a non-convex result for models which have a non-convex bare potential (like the $N=1$ linear sigma model). Also this effective potential is complex in certain regions. This seems to contradict the general proof that an effective potential is always convex and real. As soon as one realizes that this proof originates in the path-integral approach, the problem is resolved. Indeed, when computing the path integral, and including \emph{all} minima of the bare potential, one finds a real and convex effective potential. (See e.g.\ \cite{Fujimoto, Bender, Cooper, vanKessel, vanKesselthesis}.) Thus the convexity problem is only apparent, and is resolved by realizing that canonical and path-integral results are \emph{different}. There is one subtlety here: the canonical and path-integral result coincide if only one minimum is taken into account in the path integral; but this seems to imply that the path-integral result can also lead to non-convex potentials. In fact, the convexity theorem is essentially non-perturbative, and a perturbative approximation to the path integral may well be non-convex, especially since the convexity breaks down together with the perturbation series.

Let us now turn to the case of Minkowski space-time. In the canonical approach the effective potential is readily computed, and in fact coincides with the Euclidean one. This is not unreasonable since the effective potential is defined at vanishing external momenta. On the other hand, it is therefore neither everywhere real nor convex. It therefore becomes important to investigate whether the path-integral approach can cure these defects as it does in the Euclidean case.

In moving the path integral from the Euclidean to the Minkowski realm, it must be realized that we need to reappraise the most important contributions to it. In the Euclidean case we are dealing with a simple probability density, and it is trivially seen that the \emph{minima} of the action furnish the largest contribution, while the \emph{maxima} actually give the smallest possible contribution. On the basis of this argument, that is already valid at the tree level, one always concentrates on the minima alone.

In the Minkowski case things are drastically different. Each path now contributes a complex phase, which not only invalidates the convexity
argument as given in \cite{Riversbook, Haymaker, Callaway, vanKessel, vanKesselthesis}, but also implies that the relevance of certain path configurations can only be argued from constructive interference between neighboring
paths. This is of course well known: the \emph{stationary} paths are the most important. However, this argument applies, at the tree level, to minima and maxima alike, and there would appear to be no reason to disregard the maxima of the potential. We shall investigate this question in detail, and show that any maxima {\em can}, in fact, be safely neglected but only because of ({\bf{a}}) loop effects, and ({\bf{b}}) the presence of  the $i\eps$-prescription. This last fact may come as a surprise but it ought to be realized that the  $i\eps$-prescription appears precisely during the Wick transition from Euclidean to Minkowski space-time in order to regulate the path integral.

In this paper we shall also investigate the $N=1$ linear sigma model (LSM) in Minkowski space-time using the path integral approach, and compute the effective potential in perturbation theory. This enables us to conclude to what extent this effective potential is real and convex.

\section{Which Extrema Contribute}\label{sectmax}

In general QFT calculations one never considers what happens around the maximum. Here we will see why and show that the generating functional, when perturbatively expanded around a maximum, goes to zero in the limit of the space-time volume $\Omega\rightarrow\infty$. To this end we will first expand the generating functional around a general extremum and then consider what happens in particular around a maximum.

A generating functional for one scalar field generically looks like:
\bq
Z = \int\mathcal{D}\f \; \exp\left({i\over\hbar}\int d^dx \left( \h\left(\partial_{\mu}\f(x)\right)^2 - V(\f(x)) + i\eps\f^2(x) \right)\right)
\eq
Here we included an $i\eps$-term to make the path integral well defined. If we also include counter terms the generating functional generically looks like:
\bq
Z = \int\mathcal{D}\f \; \exp\left({i\over\hbar}\int d^dx \left( \h\left(\partial_{\mu}\f(x)\right)^2 + \h\delta_Z\left(\partial_{\mu}\f(x)\right)^2 - V(\f(x)) - \delta V(\f(x)) + i\eps\f^2(x) \right)\right)
\eq
Here $\delta V$ includes the mass and coupling constant counter terms.

Now we will expand the action around an extremum $\f_{\mathrm{e}}$ of the bare potential $V$. Defining
\bq
\f(x) \equiv \f_{\mathrm{e}} + \eta(x) \;, 
\eq
we get:
\bqa
Z &=& \int\mathcal{D}\eta \; \exp\Bigg({i\over\hbar}\int d^dx \Bigg( \h\left(\partial_{\mu}\eta(x)\right)^2 + \h\delta_Z\left(\partial_{\mu}\eta(x)\right)^2 + \nonumber\\
& & \phantom{\int\mathcal{D}\eta \; \exp\Bigg({i\over\hbar}\int d^dx \Bigg(} -V(\f_{\mathrm{e}}) - \h\frac{d^2V}{d\f^2}(\f_{\mathrm{e}}) \; \eta^2(x) - {1\over6}\frac{d^3V}{d\f^3}(\f_{\mathrm{e}}) \; \eta^3(x) + \ldots + \nonumber\\
& & \phantom{\int\mathcal{D}\eta \; \exp\Bigg({i\over\hbar}\int d^dx \Bigg(} -\delta V(\f_{\mathrm{e}}) - \frac{d\delta V}{d\f}(\f_{\mathrm{e}}) \; \eta(x) -\h\frac{d^2\delta V}{d\f^2}(\f_{\mathrm{e}}) \; \eta^2(x) + \ldots + \nonumber\\
& & \phantom{\int\mathcal{D}\eta \; \exp\Bigg({i\over\hbar}\int d^dx \Bigg(} i\eps\f_{\mathrm{e}}^2 + 2i\eps\f_{\mathrm{e}}\eta(x) + i\eps\eta^2(x) \Bigg)\Bigg)
\eqa
If we are dealing with a maximum of $V$ we have that
\bq \label{casemax}
\frac{d^2V}{d\f^2}(\f_{\mathrm{e}}) \equiv M^2 < 0 \;,
\eq
whereas for a minimum we have that
\bq
\frac{d^2V}{d\f^2}(\f_{\mathrm{e}}) \equiv M^2 > 0 \;.
\eq

In this paper we will only consider one loop effects, so we will keep only the Gaussian terms in the generating functional. This means we discard all interaction terms, and all Green's functions computed from our generating functional will be correct up to one loop. In this approximation the generating functional around one extremum becomes:
\bqa
Z &\approx& \exp\left({i\Omega\over\hbar}\left(-V(\f_{\mathrm{e}})-\delta V(\f_{\mathrm{e}})+i\eps\f_{\mathrm{e}}^2\right)\right) \cdot \nonumber\\
& & \int\mathcal{D}\eta \; \exp\left({i\over\hbar}\int d^dx \left( \h\left(\partial_{\mu}\eta(x)\right)^2 - \h\left(M^2-2i\eps\right) \eta^2(x) + 2i\eps\f_{\mathrm{e}}\eta(x) \right)\right)
\eqa
The path integral in the second line can be performed by first completing the square to get rid of the $\eta$-term and then shifting the $\eta$-field by 
\bq
\frac{i\eps\f_{\mathrm{e}}}{\h M^2-i\eps} \;.
\eq
Notice that this is a complex shift, such that the $\eta$-integrals now run through the complex plane. However, because the function integrated over is analytic and vanishes at $\eta=\pm\infty$ (because of the $i\eps$-term) we can bring the $\eta$-integrals back to normal integrals over the real line. In this way the generating functional becomes:
\bqa
Z &\approx& \exp\left(-{i\Omega\over\hbar}V(\f_{\mathrm{e}})-{i\Omega\over\hbar}\delta V(\f_{\mathrm{e}})-{\Omega\eps\over\hbar}\f_{\mathrm{e}}^2 - {i\Omega\over\hbar}\frac{\eps^2\f_{\mathrm{e}}^2}{\h M^2-i\eps}\right) \cdot \nonumber\\
& & \int\mathcal{D}\eta \; \exp\left({i\over\hbar}\int d^dx \left( \h\left(\partial_{\mu}\eta(x)\right)^2 - \h\left(M^2-2i\eps\right) \eta^2(x) \right)\right)
\eqa
Using
\bq \label{Gaussianpathint}
\int \mathcal{D}\eta \; \exp\left({i\over\hbar} \int d^dx \left( \h\left(\partial_{\mu}\eta\right)^2 - \h M^2\eta^2 \right)\right) \sim \exp\left(-\h\Omega {1\over(2\pi)^d}\int d^dk \; \ln\left(k^2-M^2\right)\right)
\eq
we find finally:
\bqa
Z &\approx& \exp\left(-{i\Omega\over\hbar}V(\f_{\mathrm{e}})-{i\Omega\over\hbar}\delta V(\f_{\mathrm{e}})-{\Omega\eps\over\hbar}\f_{\mathrm{e}}^2 - {i\Omega\over\hbar}\frac{\eps^2\f_{\mathrm{e}}^2}{\h M^2-i\eps}\right) \cdot \nonumber\\
& & \qquad \exp\left(-\h\Omega{1\over(2\pi)^d}\int d^dk \; \ln\left(k^2-M^2+2i\eps\right)\right) \label{genfunctgen}
\eqa

\subsection{The Maximum}

Now we will consider what happens in the case (\ref{casemax}). The integral in (\ref{genfunctgen}), after doing a Wick rotation becomes:
\bq
I \equiv {1\over(2\pi)^d}\int d^dk \; \ln\left(k^2-M^2+2i\eps\right) = {i\over(2\pi)^d}\int d^dk_E \; \ln\left(-k_E^2-M^2+2i\eps\right)
\eq
To evaluate it we say:
\bqa
\frac{\partial}{\partial M} I &=& -2M{i\over(2\pi)^d}\int d^dk_E \; \frac{1}{-k_E^2-M^2+2i\eps} \nonumber\\
&=& 2M{i\over(2\pi)^d}\frac{\pi^{d/2}}{\Gamma(d/2)} \int_0^\infty dt \; \frac{t^{d/2-1}}{t+M^2-2i\eps}
\eqa
To evaluate the $t$-integral we use integral 3.194, 3 from \cite{Gradshteyn}. Notice that in order to use this formula we have to choose
\bq
\left|\arg\left(M^2-2i\eps\right)\right| < \pi \;.
\eq
This will become very important to see what the imaginary part of the $t$-integral is, which will in turn decide whether generating functional around a maximum goes to zero or not.

We find:
\bqa
\frac{\partial}{\partial M} I &=& 2M{i\over(4\pi)^{d/2}}\Gamma(1-d/2)\left(M^2-2i\eps\right)^{d/2-1} \nonumber\\
I &=& -{i\over(4\pi)^{d/2}}\Gamma(-d/2)\left(M^2-2i\eps\right)^{d/2} \label{Iint}
\eqa
Now we are interested in the real part of this expression. This can be worked out explicitly for any $d$, from 1 to 4. Remember that for $d>4$ the theory is not renormalizable anymore, so we will not consider these cases. Notice that for the odd dimensions (\ref{Iint}) contains no singularities, so we can just put $d$ to 1 or 3. For the even dimensions it does. These singularities are always purely imaginary and thus uninteresting for our purpose. Also these singularities are of course cancelled by the counter terms. So for dimensions 2 and 4 the second term in the expansion around $d=2$ or 4 is important.

In this way one can see that $I$ is purely imaginary when $M^2>0$. Also one can see that the real part of $I$, for the case $M^2<0$, is always \emph{positive}. This means that for a maximum we have that:
\bqa
\lim_{\eps\rightarrow0} Z &\approx& \exp\left(-{i\Omega\over\hbar}V(\f_{\mathrm{e}})-{i\Omega\over\hbar}\delta V(\f_{\mathrm{e}})\right) \exp\left(-\h\Omega I\right) \nonumber\\
Z &\overset{\Omega\rightarrow\infty}{\longrightarrow}& 0
\eqa
Thus the maximum \emph{never} gives a contribution to the complete path integral.

Notice that this proof is quite general, as soon as on has an extremum of the potential where one direction has a negative second derivative (i.e.\ negative mass) the contribution of this extremum goes to zero because of the quantum fluctuations and the $i\eps$-prescription.

\section{The $N=1$ LSM}

Now we will consider the $N=1$ LSM and calculate the effective potential and some Green's functions.

The bare action of the $N=1$ LSM, including a source term, is:
\bq \label{actionws}
S = \int d^dx \; \left( \h\left(\partial_{\mu}\f\right)^2 + \h\mu\f^2 - {\lambda\over24}\f^4 + J\f + i\eps\f^2 \right) \;,
\eq
where $\lambda>0$ and $\mu>0$, to have a non-convex bare potential. We limit ourselves to the case where $J$ is constant over space-time, since we are only interested in the effective \emph{potential} and not the complete effective action. Also notice that we included an $i\eps$-term in the action again, to make the path integral well defined.

\subsection{The Effective Potential} \label{effpotsect}

To calculate the effective potential we first need the renormalized generating functional $Z(J)$. So first we introduce renormalized quantities in the usual way:
\bqa
\f^{\mathrm{R}} &\equiv& {1\over\sqrt{Z}} \f \;, \quad Z \equiv 1 + \delta_Z \nonumber\\
\mu^{\mathrm{R}} &\equiv& \mu Z - \delta_{\mu} \nonumber\\
\lambda^{\mathrm{R}} &\equiv& \lambda Z^2 - \delta_{\lambda} \label{renquant}
\eqa
The source $J$ is renormalized as:
\bq
J^{\mathrm{R}} \equiv \sqrt{Z} J
\eq
By renormalizing $J$ in this way we ensure that by taking derivatives with respect to $J^{\mathrm{R}}$ we obtain the renormalized Green's functions. From here on we will drop the superscript R, understanding that we always work with renormalized quantities. The action becomes: 
\bqa
S &=& \int d^dx \; \bigg( \h\left(\partial_{\mu}\f\right)^2 + \h\mu\f^2 - {\lambda\over24}\f^4 + J\f + i\eps\f^2 \nonumber\\[5pt]
& & \phantom{\int d^dx \; \bigg(} \h\delta_Z\left(\partial_{\mu}\f\right)^2 + \h\delta_{\mu}\f^2 - {\delta_{\lambda}\over24}\f^4 \bigg) \;. \label{renactionws}
\eqa
Notice that, in principle the $\eps$ in here is different than the $\eps$ in (\ref{actionws}) because of the renormalization, however since $\eps$ will go to zero at the end anyway we will not write this difference explicitly.

Looking at the first line of (\ref{renactionws}) (excluding the $i\eps$-term) we notice we have to distinguish between three cases: $J>\frac{2\mu v}{3\sqrt{3}}$, $-\frac{2\mu v}{3\sqrt{3}}<J<\frac{2\mu v}{3\sqrt{3}}$ and $J<-\frac{2\mu v}{3\sqrt{3}}$. Here $v$ is given by:
\bq
v \equiv \sqrt{{6\mu\over\lambda}}
\eq
For the first and last case the bare potential has one minimum, for the second case it has two minima and a maximum.

\subsubsection{$J>\frac{2\mu v}{3\sqrt{3}}$}

For this case the bare potential has only \emph{one} extremum, which is a minimum. This minimum, which we denote by $\f_+$ satisfies
\bq
-\mu\f_+ + {\lambda\over6}\f_+^3 - J = 0 \;,
\eq
and is given explicitly by
\bq
\f_+ = {v\over6} \frac{\left(108{J\over\mu v}+12\sqrt{81{J^2\over\mu^2v^2}-12}\right)^{2/3} + 12}{\left(108{J\over\mu v}+12\sqrt{81{J^2\over\mu^2v^2}-12}\right)^{1/3}} \;.
\eq

Now the generating functional is given by:
\bq
Z(J) = \int\mathcal{D}\f \; \exp\left({i\over\hbar} S(\f)\right)
\eq
Introducing the field $\eta$, which indicates the deviation of the $\f$-field from the minimum $\f_+$,
\bq
\f(x) \equiv \f_+ + \eta(x) \;,
\eq
we can use the general result (\ref{genfunctgen}) to compute the generating functional around one specific extremum in the case of the $N=1$ LSM. For the bare potential $V$ and the counter terms in (\ref{genfunctgen}) we use of course:
\bqa
V(\f) &=& V_0(\f) - J\f \nonumber\\
\delta V(\f) &=& -\h\delta_{\mu}\f^2 + {\delta_{\lambda}\over24}\f^4 \;,
\eqa
with
\bq
V_0 = -\h\mu\f^2 + {\lambda\over24}\f^4 \;.
\eq
The generating functional becomes:
\bqa
Z(J) &\approx& \exp\left({i\over\hbar}\Omega\left(J\f_+ - V_0(\f_+) + i\eps\f_+^2 + \h\delta_\mu\f_+^2 - {\delta_\lambda\over24}\f_+^4 - \frac{\eps^2\f_+^2}{\h\mu\left(3\f_+^2/v^2-1\right)-i\eps} \right)\right) \cdot \nonumber\\
& & \qquad \exp\left(-\h\Omega{1\over(2\pi)^d}\int d^dk \; \ln\left[ k^2 - \mu\left(3{\f_+^2\over v^2}-1\right) + 2i\eps \right] \right)
\eqa

The counter terms in the first line and the whole second line we recognize as the one-loop effective potential $V_1$ from the canonical approach. (For formulas giving this one-loop effective potential see \cite{Peskin, vanKesselthesis}.) So finally our result becomes:
\bq
Z(J) \approx \exp\left({i\over\hbar}\Omega\left(J\f_+ - V_0(\f_+) - V_1(\f_+) + i\eps\f_+^2 - \frac{\eps^2\f_+^2}{\h\mu\left(3\f_+^2/v^2-1\right)-i\eps} \right)\right)
\eq

Now we can obtain the $\f$-tadpole for constant source:
\bq
\langle \f \rangle(J) = -i\hbar{1\over\Omega}\frac{\partial}{\partial J} \ln Z(J) \approx \f_+ - \frac{\partial V_1(\f_+)}{\partial\f_+}\frac{\partial\f_+}{\partial J} + \mathcal{O}(\eps) \qquad \textrm{for $J>\frac{2\mu v}{3\sqrt{3}}$}
\eq
This is just the canonical result, which is expected to hold true if there is only one minimum.

Of course the treatment for the case $J<-\frac{2\mu v}{3\sqrt{3}}$ is completely similar to the case treated here, also in this case there is \emph{one} minimum, $\f_-$, and the result for the tadpole is:
\bq
\langle \f \rangle(J) \approx \f_- - \frac{\partial V_1(\f_-)}{\partial\f_-}\frac{\partial\f_-}{\partial J} + \mathcal{O}(\eps) \qquad \textrm{for $J<-\frac{2\mu v}{3\sqrt{3}}$}
\eq
with
\bq
\f_- = -{v\over6} \frac{\left(-108{J\over\mu v}+12\sqrt{81{J^2\over\mu^2v^2}-12}\right)^{2/3} + 12}{\left(-108{J\over\mu v}+12\sqrt{81{J^2\over\mu^2v^2}-12}\right)^{1/3}} \;.
\eq

\subsubsection{$-\frac{2\mu v}{3\sqrt{3}}<J<\frac{2\mu v}{3\sqrt{3}}$}

In this interesting region we have three extrema of the first line of (\ref{renactionws}) (excluding the $i\eps$-term). Two are minima and one is a maximum. In this region for $J$ it is convenient to parameterize the constant source $J$ as follows:
\bq
J = {2\mu v\over3\sqrt{3}} \sin(3\alpha) \;, \quad -{\pi\over6} < \alpha \leq {\pi\over6}
\eq
Now the two minima of the first line in (\ref{renactionws}) are given by:
\bq
\f_\pm = {2v\over\sqrt{3}} \sin\left(\alpha\pm{\pi\over3}\right)
\eq
The maximum is given by:
\bq
\f_0 = -{2v\over\sqrt{3}} \sin\alpha
\eq

If we assume that the extrema do not communicate, a good approximation for the generating functional $Z(J)$ is given by
\bq
Z(J) \approx Z_+(J) + Z_-(J) + Z_0(J)
\eq
where $Z_+$, $Z_-$ and $Z_0$ are respectively the Gaussian approximations to the generating functional around the plus-minimum, the minus-minimum and the maximum.

Now we know from section \ref{sectmax} that the maximum \emph{never} gives a contribution, i.e.\ 
\bq
Z_0 \overset{\Omega\rightarrow\infty}{\longrightarrow} 0 \;,
\eq
so we can discard $Z_0$ immediately from the formula above.

The generating functionals $Z_+$ and $Z_-$ can be calculated in the same way as we did for the case $J>\frac{2\mu v}{3\sqrt{3}}$. We find:
\bq \label{genfunctallextr}
Z_{\pm}(J) \approx \exp\left({i\over\hbar}\Omega\left(J\f_{\pm} - V_0(\f_{\pm}) - V_1(\f_{\pm}) + i\eps\f_{\pm}^2 - \frac{\eps^2\f_{\pm}^2}{\h\mu\left(3\f_{\pm}^2/v^2-1\right)-i\eps} \right)\right)
\eq

We know that the canonical one-loop effective potential $V_1(\f)$ becomes complex for $\f^2<{v^2\over3}$, everywhere else it is real \cite{Peskin, vanKesselthesis}. So $V_1(\f_\pm)$ is always real. In $Z_0$, $V_1(\f_0)$ would be complex, indicating also that $Z_0$ goes to zero as $\Omega\rightarrow\infty$, as was demonstrated explicitly in section \ref{sectmax}.

Notice that $Z_\pm$ are both merely phase factors (if we ignore the small $i\eps$-terms). This means that in the whole interval $-\frac{2\mu v}{3\sqrt{3}}<J<\frac{2\mu v}{3\sqrt{3}}$ \emph{both} minima contribute. This is completely different from the Euclidean case, where, for large $\Omega$, $Z_+$ dominates for $J>0$ and $Z_-$ dominates for $J<0$.

Also at higher than Gaussian (i.e.\ one-loop) order $Z_\pm$ will remain phase factors. One can see this by writing down the $n$-loop vacuum bubbles for $Z_\pm(J)$ for constant source $J$, which are purely imaginary. These vacuum bubbles then exponentiate and one is left with a pure phase. Thus the physics does not change fundamentally at higher order.

Notice also that, because both $Z_+$ and $Z_-$ can be written in terms of the canonical effective potential, the same set of (canonical) counter terms suffice to make the generating functional and Green's functions finite. This is of course a necessary condition for the path-integral results to make sense at all.

To calculate $Z$ further define:
\bqa
& & \alpha \equiv \h(\f_++\f_-) \;, \quad \beta \equiv \h(\f_+-\f_-) \;, \nonumber\\
& & \delta \equiv \h(f(\f_+)+f(\f_-)) \;, \quad \gamma \equiv \h(f(\f_+)-f(\f_-)) \;, \nonumber\\
& & A_0 \equiv \h\left(V_0(\f_+)+V_0(\f_-)\right) \;, \quad A_1 \equiv \h\left(V_1(\f_+)+V_1(\f_-)\right) \;, \nonumber\\
& & B_0 \equiv \h\left(V_0(\f_+)-V_0(\f_-)\right) \;, \quad B_1 \equiv \h\left(V_1(\f_+)-V_1(\f_-)\right) \;,
\eqa
with
\bq
f(\f) \equiv \frac{\h\mu i\eps\f^2\left(3\f^2/v^2-1\right)}{\h\mu\left(3\f^2/v^2-1\right)-i\eps} \;.
\eq
Then we can write:
\bq
Z_\pm(J) = \exp\left({i\over\hbar}\Omega\left(J(\alpha\pm\beta) - (A_0\pm B_0) - (A_1\pm B_1) + (\delta\pm\gamma) \right)\right)
\eq
\bq
Z(J) \approx 2\exp\left({i\over\hbar}\Omega\left(J\alpha-A_0-A_1+\delta\right)\right) \cos\left({\Omega\over\hbar}\left(J\beta-B_0-B_1+\gamma\right)\right)
\eq
For the $\f$-tadpole we find:
\bqa
\langle\f\rangle(J) &\approx& -i\hbar{1\over\Omega}\frac{\partial}{\partial J} \ln Z \nonumber\\
&=& \alpha - \frac{\partial A_1}{\partial J} + \frac{\partial\delta}{\partial J} + i\left(\beta-\frac{\partial B_1}{\partial J}+\frac{\partial\gamma}{\partial J}\right) \tan\left({\Omega\over\hbar}(J\beta-B_0-B_1+\gamma)\right) \nonumber\\
&=& \alpha - \frac{\partial A_1}{\partial J} + i\left(\beta-\frac{\partial B_1}{\partial J}\right) \tan\left({\Omega\over\hbar}(J\beta-B_0-B_1)\right) + \mathcal{O}(\eps) \qquad \textrm{for $-\frac{2\mu v}{3\sqrt{3}}<J<\frac{2\mu v}{3\sqrt{3}}$} \nonumber\\ \label{phitadpole}
\eqa

\begin{figure}[h]
\begin{center}
\epsfig{file=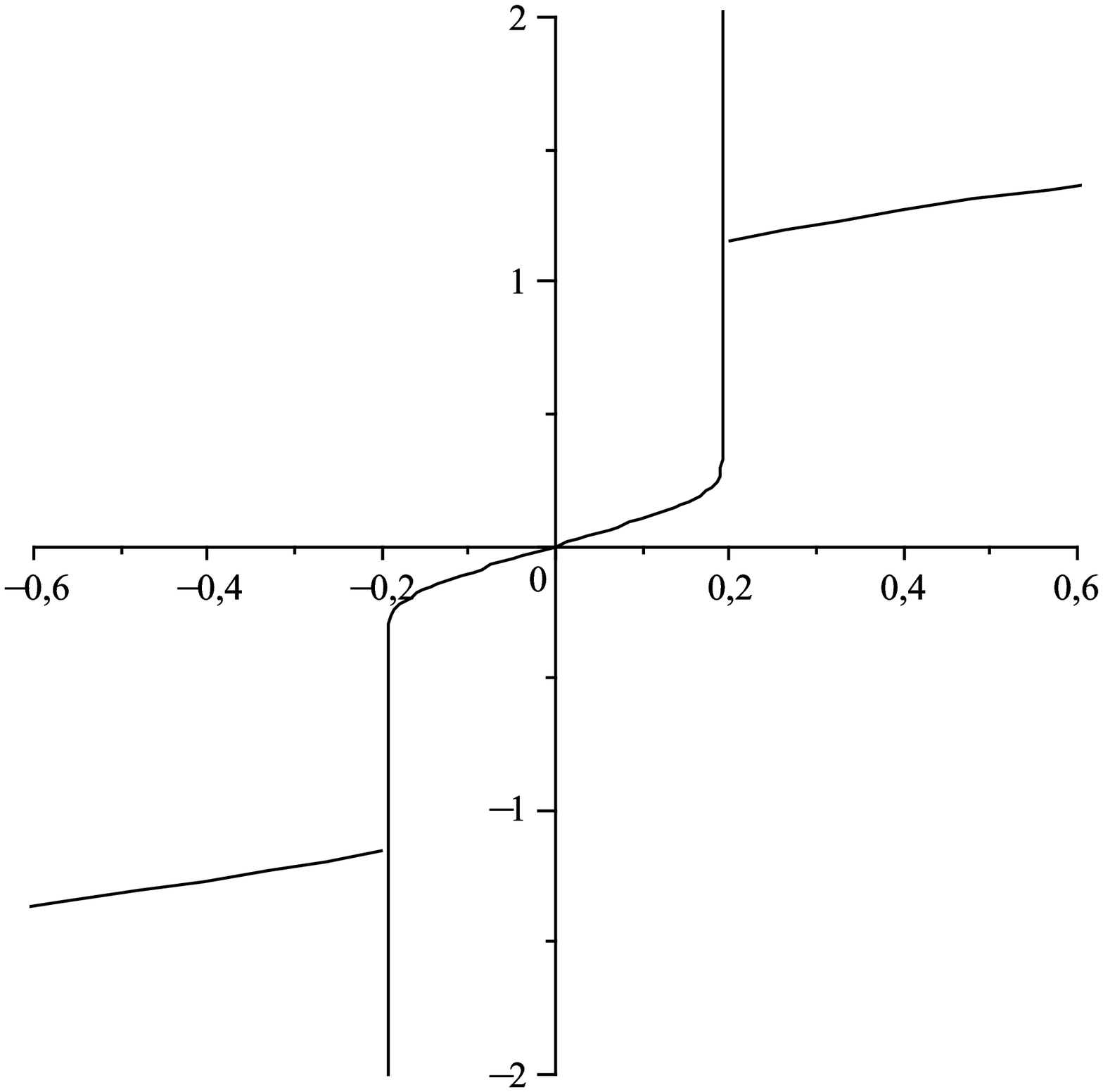,width=7cm}
\end{center}
\caption{The real part of the $\f$-tadpole as a function of $J$ for $\hbar=0.01$, $\mu=\h$, $v=1$, $\Omega=100$ and dimension $d=1$.}
\label{figRephid1}
\end{figure}

\begin{figure}[h]
\begin{center}
\epsfig{file=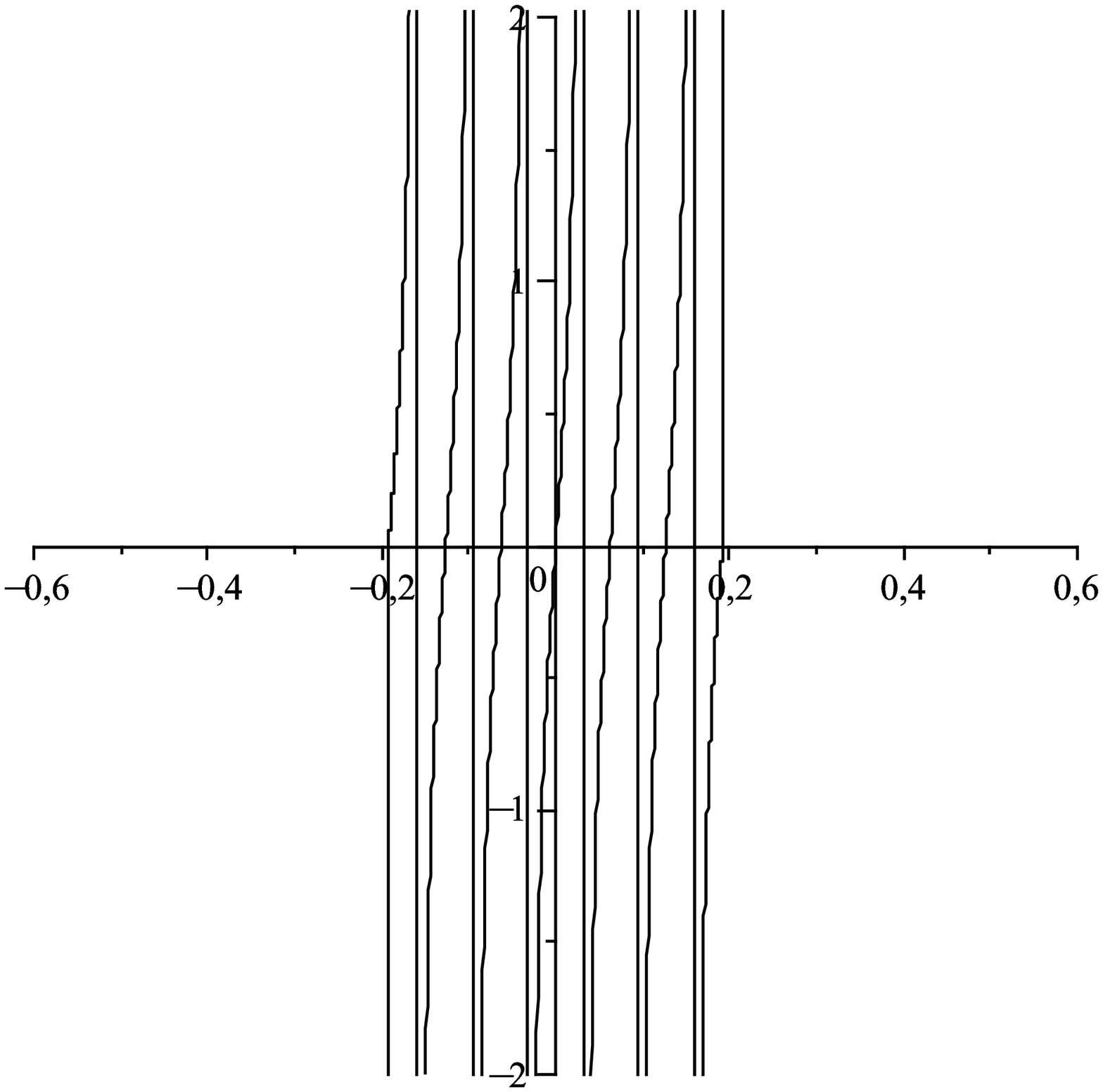,width=7cm}
\end{center}
\caption{The imaginary part of the $\f$-tadpole as a function of $J$ for $\hbar=0.1$, $\mu=\h$, $v=1$, $\Omega=5$ and dimension $d=1$.}
\label{figImphid1}
\end{figure}

\begin{figure}[h]
\begin{center}
\epsfig{file=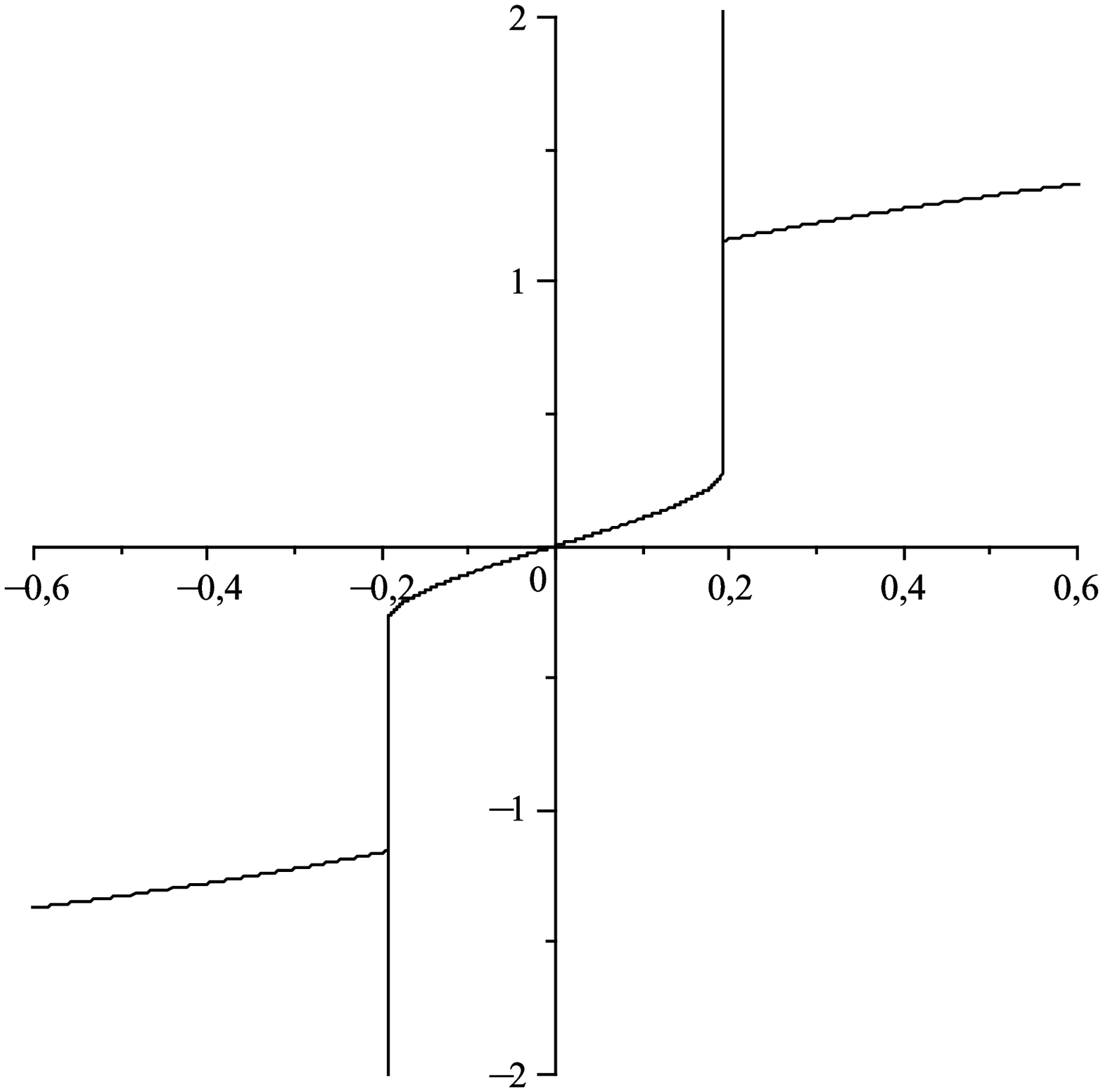,width=7cm}
\end{center}
\caption{The real part of the $\f$-tadpole as a function of $J$ for $\hbar=0.01$, $\mu=\h$, $v=1$, $\Omega=100$ and dimension $d=4$.}
\label{figRephid4}
\end{figure}

\begin{figure}[h]
\begin{center}
\epsfig{file=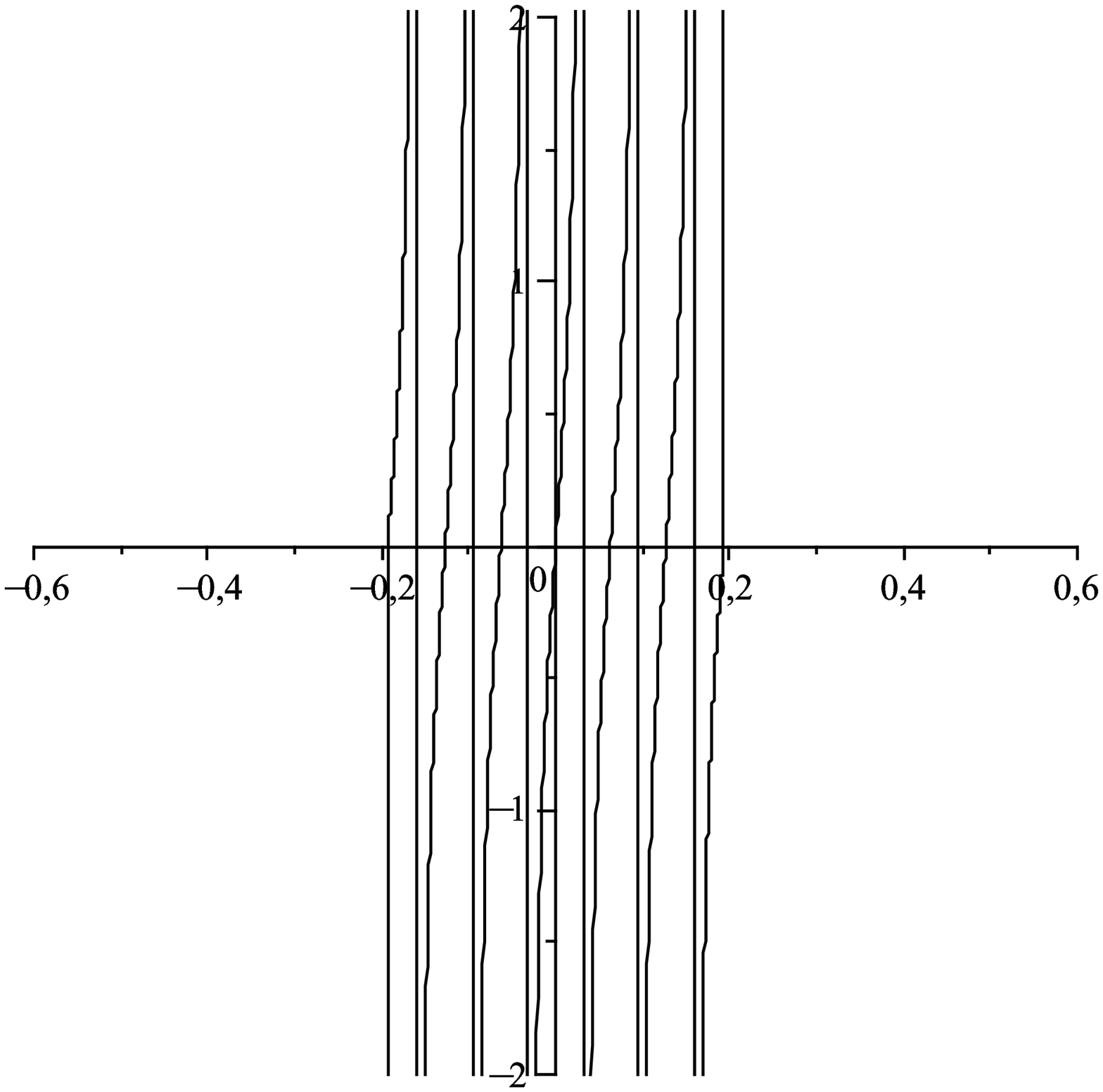,width=7cm}
\end{center}
\caption{The imaginary part of the $\f$-tadpole as a function of $J$ for $\hbar=0.1$, $\mu=\h$, $v=1$, $\Omega=5$ and dimension $d=4$.}
\label{figImphid4}
\end{figure}

This result for the $\f$-tadpole is plotted in figures \ref{figRephid1} and \ref{figImphid1}, for all $J$-regions, and for dimension $d=1$. In figures \ref{figRephid4} and \ref{figImphid4} the $\f$-tadpole is plotted for $d=4$.

First notice the singularities in the tadpole at the $J$-values $J=\pm{2\mu v\over3\sqrt{3}}$. Both the real and imaginary part of the result for the case $-{2\mu v\over3\sqrt{3}}<J<{2\mu v\over3\sqrt{3}}$ blow up at the boundaries. This is understood by realizing that, for example at $J=+{2\mu v\over3\sqrt{3}}$, the derivative of $V_1(\f_-)$ with respect to $J$ blows up because $\frac{\partial\f_-}{\partial J}$ blows up there. Physically one can say that, at $J=+{2\mu v\over3\sqrt{3}}$, the minimum becomes so unstable that perturbation theory can no longer be used. Thus what happens exactly at the points $J=\pm{2\mu v\over3\sqrt{3}}$ cannot be understood within perturbation theory. The singularities we see might even be artifacts of perturbation theory, however without exactly calculating the path integral at this $J$-value one cannot resolve this problem, and we shall not go into this further.

Secondly notice that the $\f$-tadpole contains a heavily oscillating imaginary part. The larger $\Omega$ becomes, the faster the tangent in (\ref{phitadpole}) will oscillate, and the closer the singular points will lie together. One might say that this imaginary part has to be interpreted as a distribution, i.e.\ that it has to be weighed with a test function. In that case the imaginary part vanishes as $\Omega\rightarrow\infty$, and the $\f$-tadpole is purely real.

Finally we can say that if the singularities at the $J$-values $J=\pm{2\mu v\over3\sqrt{3}}$ are artifacts of perturbation theory, and the actual result gives a nice interpolation between $J$-values just below and above $J=\pm{2\mu v\over3\sqrt{3}}$, \emph{and} if the imaginary part of the tadpole indeed goes to zero as $\Omega\rightarrow\infty$, then we have a purely real and monotonically increasing result for the $\f$-tadpole. In that case the effective potential is also real and convex. However, on the other hand, if these two conditions are not satisfied, the effective potential is not real and convex, it might not even exist perturbatively. This is not as strange as it sounds. Of course all Green's functions exist, and one might argue that the effective potential is related to a sum over all 1PI Green's functions, and therefore must also exist. However, it is not clear whether the sum of all 1PI Green's functions is a convergent one in case of the path-integral approach.

\subsection{The Green's Functions}

Now we calculate the Green's functions of the model, also taking into account both minima (and not the maximum) and making the Gaussian approximation.

First we compute the tadpole:
\bqa
\langle \f(x) \rangle &=& \frac{\int \mathcal{D}\f \; \f(x) \; e^{{i\over\hbar}S(J=0)}}{\int \mathcal{D}\f \; e^{{i\over\hbar}S(J=0)}} \nonumber\\
&\approx& \frac{(\f_+(0)+\langle\eta(x)\rangle_+(0)) Z_+(0) + (\f_-(0)+\langle\eta(x)\rangle_-(0)) Z_-(0)}{Z_+(0) + Z_-(0)}
\eqa
Here $\langle\eta\rangle_\pm(J)$ is given by:
\bq
\f_\pm(J) + \langle\eta\rangle_\pm(J) = -{i\hbar\over\Omega}{\partial\over\partial J} \ln Z_\pm(J) \;.
\eq
This is just the $\eta$-tadpole from the canonical approach. Now we know:
\bqa
\f_\pm(0) &=& \pm v \nonumber\\
Z_\pm(0) &=& Z_+ \nonumber\\
\langle\eta\rangle_\pm(0) &=& \pm\langle\eta\rangle_+(0)
\eqa
So we find
\bq
\langle \f(x) \rangle = 0 \;.
\eq

In the same way we can find the two-point Green's function.
\bqa
\langle \f(x)\f(y) \rangle &=& \frac{\int \mathcal{D}\f \; \f(x)\f(y) \; e^{{i\over\hbar}S(J=0)}}{\int \mathcal{D}\f \; e^{{i\over\hbar}S(J=0)}} \nonumber\\
&\approx& \frac{\langle(\f_++\eta(x))(\f_++\eta(y))\rangle_+(0) Z_+(0) + \langle(\f_-+\eta(x))(\f_-+\eta(y))\rangle_-(0) Z_-(0)}{Z_+(0) + Z_-(0)} \nonumber\\
&=& v^2 + 2v \langle\eta\rangle_+(0) + \langle\eta(x)\eta(y)\rangle_+(0) \nonumber\\
\langle \f(x)\f(y) \rangle_{\mathrm{conn}} &\approx& v^2 + 2v \langle\eta\rangle_+(0) + \langle\eta(x)\eta(y)\rangle_+(0)
\eqa
Here $\langle\eta(x)\eta(y)\rangle_+(0)$ is again just the $\eta$-propagator from the canonical approach.

Clearly the tadpole and the connected $\f$-propagator (it is questionable whether one can still call this a propagator) are \emph{different} from the result of the canonical approach. Notice however that the \emph{even complete} Green's functions are identical in both approaches. (By complete we mean disconnected plus connected pieces.) This is due to the fact that in the $N=1$ LSM we only have two minima, summing over these minima for the \emph{even complete} Green's functions leaves them the same. In the path-integral approach the \emph{odd complete} Green's functions are zero, because of the symmetry.

Because in the path-integral approach the odd complete Green's functions are zero, and the even ones are identical to the canonical complete Green's functions a check with the Schwinger-Dyson equations is trivial. If the canonical Green's functions satisfy the S-D equations (and they do) then the path-integral Green's functions will also satisfy them, because the S-D equations are linear.

\section{Acknowledgement}

M. van Kessel acknowledges the support of the FOM network "Theoretical particle physics in the era of the LHC". E. Argyres wishes to acknowledge the support of EU project RTN MRTN-CT-2006-035505 Heptools.

\end{document}